\documentclass[%
twocolumn,
 reprint,
 amsmath,amssymb,
 aps,
]{revtex4-1}

\usepackage{graphicx}% Include figure files
\usepackage{dcolumn}% Align table columns on decimal point
\usepackage{bm}% bold math
\usepackage{hyperref}% add hypertext capabilities
\hypersetup{
	colorlinks=true,
	linkcolor=blue,%(pigment),
	filecolor=magenta,      
	urlcolor=blue,%cyan,
	citecolor=blue,%red,%(pigment),
	anchorcolor=blue,%(pigment),
}

\begin{document}

\preprint{APS/123-QED}

%\thanks{A footnote to the article title}%
\title{Fractional Quantum Hall States of the $\mathcal{A}$ phase in the Second Landau Level}
\author{Sudipto Das}
%\email{sudiptodas.cgr@gmail.com}
% \altaffiliation[Also at ]{Physics Department, XYZ University.}%Lines break automatically or can be forced with \\
\author{Sahana Das}%
%\email{sahanadas1993@gmail.com}
\author{Sudhansu S. Mandal}
%\email{sudhansu@phy.iitkgp.ac.in}
\affiliation{%
	Department of Physics, Indian Institute of Technology, Kharagpur, West Bengal 721302, India
}%

%\collaboration{MUSO Collaboration}%\noaffiliation
%
%\author{Charlie Author}
% \homepage{http://www.Second.institution.edu/~Charlie.Author}
%\affiliation{
% Second institution and/or address\\
% This line break forced% with \\
%}%
%\affiliation{
% Third institution, the second for Charlie Author
%}%
%\author{Delta Author}
%\affiliation{%
% Authors' institution and/or address\\
% This line break forced with \textbackslash\textbackslash
%}%
%
%\collaboration{CLEO Collaboration}%\noaffiliation

\date{\today}% It is always \today, today,
%  but any date may be explicitly specified

\begin{abstract}
	A proposal of the existence of an {\em Anomalous} phase ($\mathcal{A}$ phase) [\href{https://journals.aps.org/prl/abstract/10.1103/PhysRevLett.131.056202}{Das {\it et al.}, Phys. Rev. Lett. {\bf 131}, 056202 (2023)}]  at the experimental range of moderate  Landau-level-mixing strength has recently been made for $5/2$ state. We here report that the gapped $\mathcal{A}$ phase is generic to the sequence of spin-polarized fractional quantum Hall states with filling fractions $\nu = n/(nm-1)$ and $\nu = 1-n/(nm-1)$, $(n \geqslant 1,\,m\geqslant 3)$, that exhausts almost all the observed states and also predicts some states in the second Landau level for GaAs systems. Our proposed trial wavefunctions for all these states have remarkably high overlaps with the corresponding exact ground states and can support non-Abelian quasiparticle excitations with charge $e/[2(nm-1)]$. By analyzing edge modes, we predict experimentally verifiable thermal Hall conductance $2.5(\pi^2 k_B^2T/3h)$ for all the states in these sequences.
\end{abstract}

%\keywords{Suggested keywords}%Use showkeys class option if keyword
%display desired
\maketitle

% @@@@@@@@@@@@@@@@@@@@@@@@@@@@ INTRODUCTION @@@@@@@@@@@@@@@@@@@@@@@@@@@@@@
% 

Surprising discovery \cite{willett_1987_ObservationEvendenominatorQuantum} of an even-denominator fractional quantum Hall effect (FQHE) state $5/2$ and subsequent observations \cite{xia_2004_ElectronCorrelationSecond,pan_2008_ExperimentalStudiesFractional,choi_2008_ActivationGapsFractional,kumar_2010_NonconventionalOddDenominatorFractional,zhang_2012_SpinPolarization12,shingla_2018_FinitetemperatureBehaviorSecond} of other odd- and even-denominator states in the second Landau level (SLL) are fascinating for the exotic electronic correlations. Following the pioneering proposal of Moore and Read (MR) \cite{moore_1991_NonabelionsFractionalQuantum} for 5/2 state,  even-denominator FQHE states are believed to host excitations of non-Abelian quasiparticles. This occurs \cite{read_2000_PairedStatesFermions} owing to the Cooper-pairing with inherent $\mathbb{Z}_2$ symmetry amongst emergent composite fermions \cite{jain_1989_CompositefermionApproachFractional,jain_2007_CompositeFermions} arising due to the interaction between electrons in the presence of a high magnetic field. There have also been proposals in terms of particle-hole conjugate \cite{lee_2007_ParticleHoleSymmetryQuantum,levin_2007_ParticleHoleSymmetryPfaffian} of MR Pfaffian state and particle-hole symmetric \cite{son_2015_CompositeFermionDirac,zucker_2016_StabilizationParticleHolePfaffian} Pfaffian state as alternative topological orders of the 5/2 state.

The odd-denominator observed states like 7/3, 8/3, 11/5, and 14/5 are believed \cite{macdonald_1986_CollectiveExcitationsFractional,dambrumenil_1988_FractionalQuantumHall,dolev_2011_CharacterizingNeutralModes,wurstbauer_2015_GappedExcitationsUnconventional,faugno_2021_UnconventionalMathbbZParton} to behave as Abelian FQHE states. On the other hand, a generalization \cite{read_1999_PairedQuantumHall} of MR theory for 5/2 state yielding $2+n/(n+2)$ sequence of states supporting quasiparticles as $\mathbb{Z}_n$ parafermions \cite{read_1999_PairedQuantumHall} predicts non-Abelian 8/3 state. This sequence, however, does not reproduce all the observed states in the SLL. Other generalizations \cite{jolicoeur_2007_NonAbelianStatesNegative,bishara_2008_QuantumHallStates,bonderson_2008_FractionalQuantumHall} of MR theory fetch certain odd- and even-denominator non-Abelian states as well. Both the Abelian and non-Abelian FQHE states are also proposed in parton model \cite{jain_1989_IncompressibleQuantumHall,balram_2018_FractionalQuantumHall,balram_2019_PartonConstructionParticleholeconjugate}.
However, none of the different kinds of topological orders predicted by these theories have yet been confirmed in any experiment.

The validity of the proposed wavefunctions based on the above mentioned theories is mainly tested numerically \cite{balram_2018_FractionalQuantumHall,balram_2019_PartonConstructionParticleholeconjugate,rezayi_2011_BreakingParticleHoleSymmetry,zaletel_2015_InfiniteDensityMatrix,pakrouski_2015_PhaseDiagramFractional,rezayi_2017_LandauLevelMixing,simon_2020_EnergeticsPfaffianAntiPfaffian,pakrouski_2016_Enigmatic12Fractional} for the pure Coulomb interaction or at the best very low strength of the Landau-level-mixing (LLM), $\kappa$, which is quantified as the ratio of the Coulomb energy scale and the cyclotron energy scale. Because the experiments are typically performed \cite{willett_1987_ObservationEvendenominatorQuantum,xia_2004_ElectronCorrelationSecond,pan_2008_ExperimentalStudiesFractional,choi_2008_ActivationGapsFractional,kumar_2010_NonconventionalOddDenominatorFractional,zhang_2012_SpinPolarization12,shingla_2018_FinitetemperatureBehaviorSecond,dean_2008_ContrastingBehaviorFractional,dean_2008_IntrinsicGapFractional,zhang_2010_FractionalQuantumHall,pan_2001_ExperimentalEvidenceSpinpolarized,pan_2014_CompetingQuantumHall,samkharadze_2017_ObservationAnomalousDensitydependent,banerjee_2018_ObservationHalfintegerThermal,dutta_2022_DistinguishingNonabelianTopological,dutta_2022_IsolatedBallisticNonabelian} between $1$ through $12$ Tesla magnetic field for these FQHE states in GaAs systems, the role of filled and empty Landau levels also become important as $\kappa$ becomes \cite{bishara_2009_EffectLandauLevel,peterson_2013_MoreRealisticHamiltonians,sodemann_2013_LandauLevelMixing,simon_2013_LandauLevelMixing} moderately higher in the range $0.7$--$2.5$. Due to the LLM \cite{bishara_2009_EffectLandauLevel,peterson_2013_MoreRealisticHamiltonians,sodemann_2013_LandauLevelMixing,simon_2013_LandauLevelMixing}, not only the two-body pseudopotentials get renormalized, but certain three-body pseudopotentials are also emerged and consequently the particle-hole symmetry no longer becomes an exact symmetry in the SLL.
Exact diagonalization study with perturbatively obtained two-body and three-body pseudopotentials \cite{peterson_2013_MoreRealisticHamiltonians} in the spherical geometry \cite{haldane_1983_FractionalQuantizationHall} suggests \cite{das_2023_AnomalousReentrantQuantum} a topological phase transition between the conventional phase and an {\em Anomalous} phase ($\mathcal{A}$ phase) at $\kappa \sim 0.7$ for 5/2 state. The ground state at this $\mathcal{A}$ phase of 5/2 state is well described by a trial wavefunction \cite{das_2023_AnomalousReentrantQuantum}, which is nearly orthogonal to MR \cite{moore_1991_NonabelionsFractionalQuantum}, particle-hole conjugate of MR \cite{lee_2007_ParticleHoleSymmetryQuantum,levin_2007_ParticleHoleSymmetryPfaffian}, and particle-hole symmetric Pfaffian \cite{son_2015_CompositeFermionDirac,zucker_2016_StabilizationParticleHolePfaffian} wavefunctions.

% @@@@@@@@@@@@@@@@@@@ BRIEF OUTLINE OF THE PAPER @@@@@@@@@@@@@@@@@@@@@@@@@
%
Here, we show, by calculating overlaps of the exact ground states of the Hamiltonian consisting of LLM-corrected two-body and three-body pseudopotentials in spherical geometry for different values of $\kappa$, the $\mathcal{A}$ phase is generic \cite{note613} to all the observed  \cite{willett_1987_ObservationEvendenominatorQuantum,xia_2004_ElectronCorrelationSecond,pan_2008_ExperimentalStudiesFractional,choi_2008_ActivationGapsFractional,kumar_2010_NonconventionalOddDenominatorFractional,zhang_2012_SpinPolarization12,shingla_2018_FinitetemperatureBehaviorSecond} FQHE states in the SLL. 
We propose trial wavefunctions for all these FQHE states as a generalization to the recently proposed trial wavefunction \cite{das_2023_AnomalousReentrantQuantum} for 5/2 state in the $\mathcal{A}$ phase. 
These wavefunctions have remarkably high overlaps with the corresponding exact ground states and are argued to support non-Abelian quasiparticle excitations as well.
The $\mathcal{A}$ phase of all these FQHE states is found to have a positive finite charge gap in the thermodynamic limit.
The counting of low-lying edge states has been obtained by calculating entanglement spectra (ES). The quasiparticle charges corresponding to the proposed trial wavefunctions have been found in the Chern-Simons formalism \cite{wen_1992_ClassificationAbelianQuantum} for our proposed sequences of filling factors $\nu = n/(nm-1)$, $(n \geqslant 1,\,m\geqslant 3)$, and $1-\nu$ in the SLL.
We further predict that all these FQHE states will provide a unique thermal Hall conductance $2.5G_0$  ($G_0=\pi^2k_B^2T/3h$) governed by two (three) bosonic downstream edge modes for two (three) filled Landau levels and one downstream (upstream) Majorana edge mode for FQHE states $\nu$ $ (1-\nu)$.

As described in Ref.~\onlinecite{das_2023_AnomalousReentrantQuantum}, the proposed wavefunction for $\nu = 1/2$ state in the SLL is interpreted as follows. 
The composite bosons (CBs)--bound-state of an electron with one unit of flux quantum--  divide themselves into two groups, CBs in each group is noninteracting and condense while the CBs between two groups repulse each other such that they feel zeros of order two at each other's position. In this letter, we generalize this in terms of dividing each group into a number of sectors and fixing a number of zeros between a pair of CBs belonging to two groups.  
Consider $ n $ fictitious sectors with an equal number of CBs in each of the two groups. Next, each CB of any sector of one group feels $ 2(m-2) $ zeros and $ 2(m-1) $ zeros at the positions of the CBs of one of the sectors and the remaining $ (n-1) $ sectors, respectively, in the other group. This fixation of zeros leads to the formation of FQHE states in the filling factor sequence
\begin{equation}\label{eq.series}
	\nu = \frac{n}{nm-1}
\end{equation}
and its particle-hole conjugate state at $\nu =1-n/(nm-1)$ for $m \geqslant 3$ and $n \geqslant 1$. The total angular momentum of the corresponding wavefunction (dropping ubiquitous Gaussian factor \cite{laughlin_1983_AnomalousQuantumHall,jain_2007_CompositeFermions})
% $ \exp [-\prod_i^N |z_i|^2/4] $ ) 
\begin{eqnarray}\label{eq.gen_wavefun}
	&&	\Psi_{\mathcal{A}}\,(n,m)  = \prod_{i<j}^N (z_i-z_j)\, {\cal S} \left[ \prod_{1\leqslant k,l\leqslant N/(2n)}  \right. \nonumber \\
	&& \left( \prod_{\alpha =0}^{n-1} \left(z_{k+\alpha N/(2n)} -z_{l+N/2+\alpha N/(2n)}\right)^{2(m-2)} \right) \nonumber \\
	& &\left.  \left( \prod_{\alpha\neq \beta,0}^{n-1}
	\left(z_{k+\alpha N/(2n)} -z_{l+N/2+\beta N/(2n)}\right)^{2(m-1)}  \right) \right]
\end{eqnarray}
that we propose for the $\mathcal{A}$ phase becomes $M = N(\nu^{-1}N-1)/2$ in the disk-geometry implying the same `flux-shift' (by $1$) in the spherical geometry for all the FQHE states in the SLL.
Here ${\cal S}$ represents symmetrization with respect to all $N$ particle indices and the complex particle coordinates $z_j = (x_j -i\, y_j)/\ell_0$ with $\ell_0$ being the magnetic length.
The CBs form due to the attachment of one flux quantum with each electron described by the  Jastrow factor $\prod_{i<j}^N (z_i-z_j)$. The indices $\alpha$ and $\beta$ represent $n$ sectors in both the condensates of CBs. 
The wavefunction $\Psi_{\mathcal{A}}\,(n,m)$ does not vanish even if up to macroscopic $N/2$ CBs of a group coincide. In analogy to other known FQHE wavefunctions \cite{moore_1991_NonabelionsFractionalQuantum,read_1999_PairedQuantumHall} supporting non-Abelian quasiparticles having similar properties (albeit for finite number of particles), we believe that $\Psi_{\mathcal {A}}\,(n,m)$ too will support non-Abelian quasiparticles.

The observed \cite{xia_2004_ElectronCorrelationSecond,pan_2008_ExperimentalStudiesFractional,choi_2008_ActivationGapsFractional,zhang_2012_SpinPolarization12,shingla_2018_FinitetemperatureBehaviorSecond} principal sequence $\nu=1/2,\, 2/5,\, 3/8, \cdots $ corresponds to $ m=3 $ and $ n=1, \,2,\,3,\cdots$ respectively in Eq.~\eqref{eq.series}. Other observed \cite{xia_2004_ElectronCorrelationSecond,pan_2008_ExperimentalStudiesFractional,choi_2008_ActivationGapsFractional,zhang_2012_SpinPolarization12,shingla_2018_FinitetemperatureBehaviorSecond} FQHE states in the SLL are  $ \nu = 1/3 $ (also 2/3) for $ m=4$ and $n=1$,  $2/9 $ (also 7/9) for $m=5$ and $ n=2 $, and   $ 1/5 $  (also 4/5) for $m=6$ and $ n=1$.
Only $\nu=6/13$ (in between two consecutive filling factors, namely $1/2$ and $2/5$ of the principal sequence) amongst observed \cite{kumar_2010_NonconventionalOddDenominatorFractional,shingla_2018_FinitetemperatureBehaviorSecond} FQHE states does not fit in the sequence \eqref{eq.series}, just as $4/11$ state \cite{mukherjee_2014_Enigmatic11State,das_2021_UnconventionalFillingFactor} in the lowest Landau level does not fit in the principal Jain sequence \cite{jain_1989_CompositefermionApproachFractional}.

\begin{figure}[t]
	\centering
	\includegraphics[width=\linewidth]{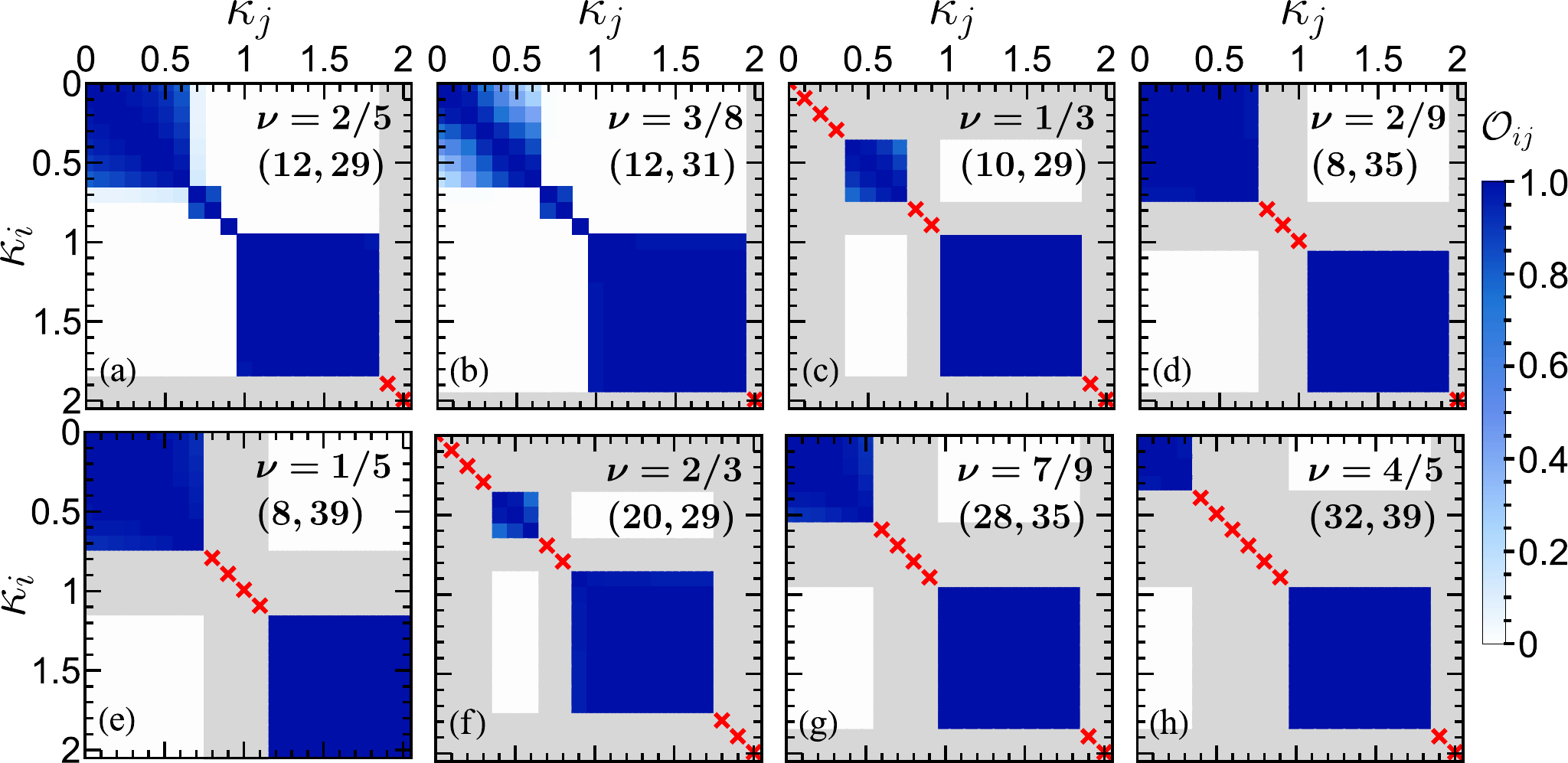}
	\caption{(color online) 
	Overlaps (shown as color map) $\mathcal{O}_{ij}$ of the exact ground states of $\hat{H}_{\text{eff}}$ in Eq.~(3) at different filling factors $\nu$ observed in the SLL with the system size mentioned as $(N, N_\Phi)$ in each of the panels. 
	No overlap has been calculated in the gray zone as one of the ground states is unquantized, i.e., the ground state is at $L\neq 0$, represented by the crossed marks. 
	}
	\label{fig.phase}
\end{figure}

%% H_Eff =======================================
The effective Hamiltonian for the spin-polarized electrons, including the effect of LLM  \cite{bishara_2009_EffectLandauLevel,peterson_2013_MoreRealisticHamiltonians} in the SLL, is given by
\begin{eqnarray}\label{eq.llmpot}
	\hat{H}_{\text{eff}}(\kappa) &=&
	\sum_{m\, \text{odd}} \left[ V_m^{(2)} + \kappa \,\delta V_m^{(2)} \right] \sum_{i<j} \hat{P}_{ij}(m) \nonumber \\
	&&	+ \sum_{m\geqslant 3}
	\kappa \, V_m^{(3)} \sum_{i<j<k} \hat{P}_{ijk}(m)
\end{eqnarray}
%for the electrons,
where, $V_m^{(2)}$ represents two-body bare Coulomb pseudopotential in the SLL and $\delta V_m^{(2)}$ is its correction due to the LLM and $V_m^{(3)}$ is the emergent three-body pseudopotential arising due to the LLM.
Here $ \hat{P}_{ij}(m) $ and $ \hat{P}_{ijk}(m) $ are two- and three-body projection operators, respectively, onto pairs or triplets of electrons with relative angular momentum $ m $.  We exactly diagonalize $\hat{H}_{\text{eff}}$ in a spherical geometry with limited pseudopotentials (see Ref.~\cite{pseudopotential}) for the observed FQHE states given by the sequences of states as shown in Eq.~\eqref{eq.series} for finite systems of $N$ electrons with the corresponding number of flux quanta $N_\Phi = \nu^{-1}N-1$. We then determine overlaps of the exact ground states when found at the total angular momentum $L=0$ for different values of $\kappa$: $\mathcal{O}_{ij} = \langle \Psi_{\text{gs}}(\kappa_i) \vert \Psi_{\text{gs}} (\kappa_j) \rangle$. As found in  Ref.~\onlinecite{das_2023_AnomalousReentrantQuantum} for $\nu =1/2$ state, we show (Fig.~\ref{fig.phase}) that $\mathcal{A}$ phase in the SLL at the moderate regime of $\kappa$ is generic to all the observed FQHE states belonging to the sequence $\nu$ and $1-\nu$ in Eq.~\eqref{eq.series}, {\em viz}, $\nu=$ 2/5, 3/8, 1/3, 2/9, 1/5, 2/3, 7/9, and 4/5. The exact ground states in the $\mathcal{A}$ phase are orthogonal to the corresponding ground states at the low-$\kappa$ phase. These two phases are intermediated by a regime of unquantized states.

\begin{figure}[]
	\centering
	\includegraphics[width=\linewidth]{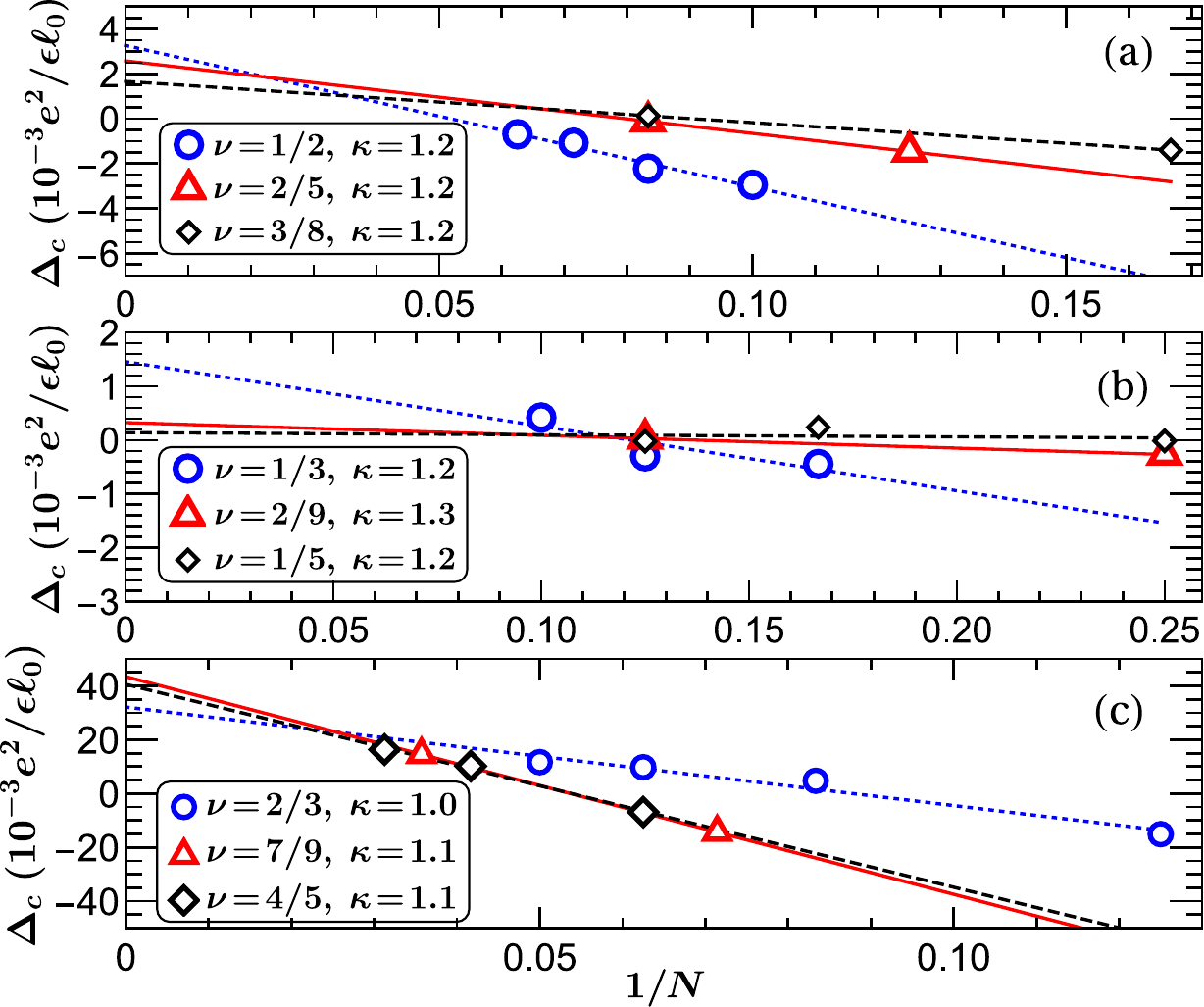}
	\caption{(color online) 
		Charge gap $\Delta_c $ for different FQHE systems scaled with $1/N$ for different $\nu$ in the $\mathcal{A}$ phase.
	}
	\label{fig.gap}
\end{figure}

%% CHARGE GAP =======================================
We calculate \cite{supplementary} charge excitation energy gap $\Delta_c$ \cite{rezayi_2021_StabilityParticleholePfaffian} for different pairs of $N$ and $N_\Phi$ for a given FQHE state by taking the average of a single quasiparticle and a single quasihole excitation energies. Fig.~\ref{fig.gap} shows the scaling of $\Delta_c$ with $1/N$ for all the observed FQHE states belonging to the sequences $\nu$ and $1-\nu$. A thermodynamic extension of this scaling indicates positive and finite excitation energies of a pair of a quasiparticle and a quasihole. Therefore, all these states in the $\mathcal{A}$ phase are quantized.

The value of charge gap shown in Fig.~\ref{fig.gap} in the thermodynamic limit is mildly sensitive \cite{supplementary} to the inclusion of three-body $V_9^{(3)}$ pseudopotential correction \cite{rezayi_2017_LandauLevelMixing}. However, this mild effect could be adverse when the magnitude of the charge gap is very small without the inclusion of $V_9^{(3)}$. This is what obtained \cite{supplementary} for 1/5 and 2/9 states as the thermodynamic gaps for these states become negative, albeit small. Therefore, a more accurate estimation of pseudopotentials (higher orders in $\kappa$) is necessary to resolve the positivity of the gap and also its increase in magnitude to rule out an outside possibility of gaplessness.

%% ES =======================================
\begin{figure}[]
	\centering
	\includegraphics[width=\linewidth]{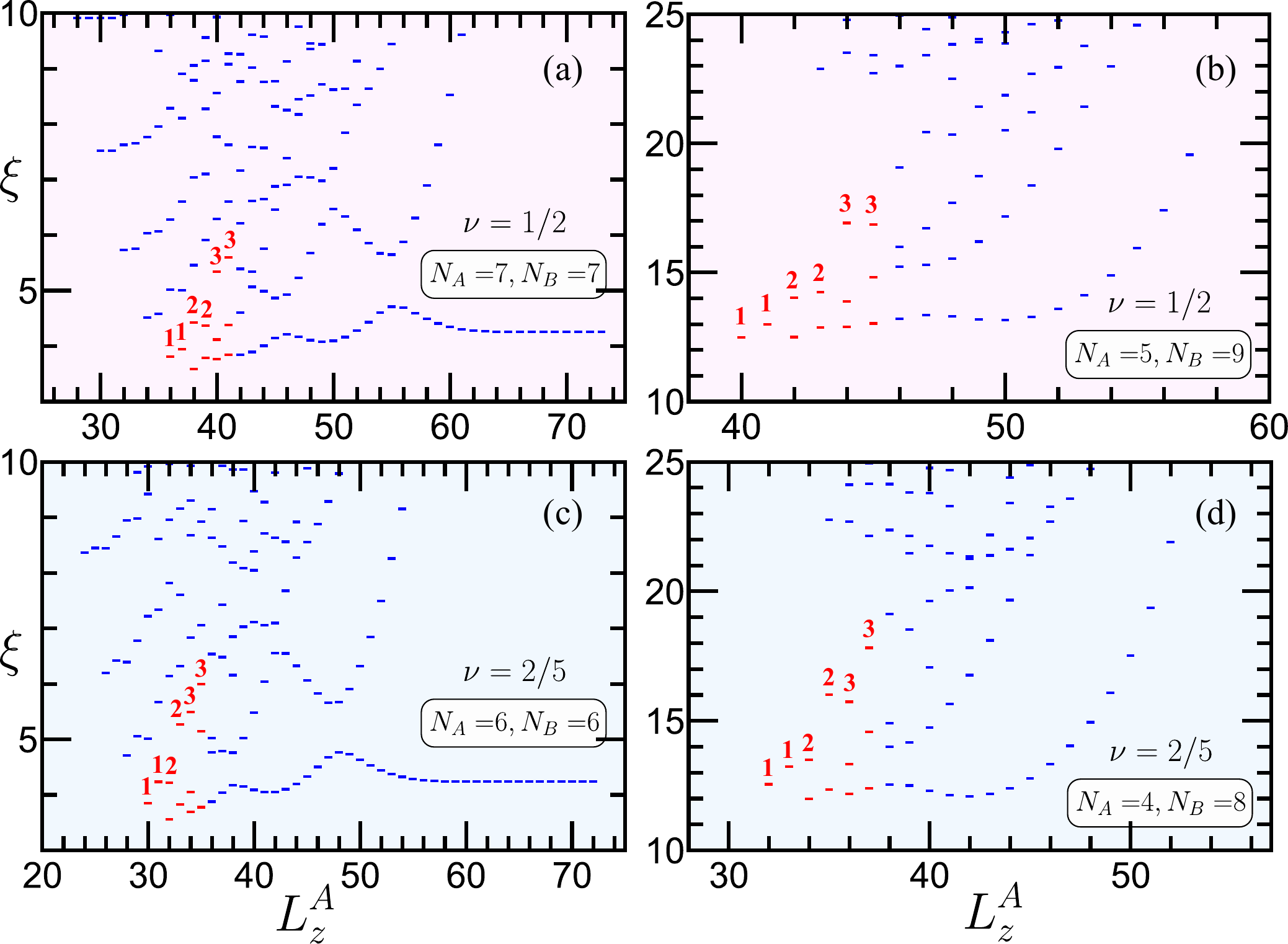}
	\caption{(color online)
		(a) ES of $ \nu=1/2 $ state in the $ \mathcal{A} $ phase ($ \kappa = 1.2 $) for flux shift of $ N_\Phi = 27 $ with an equal number of electrons $ N_A = N_B = 7 $ in both the partitions $A$ and $B$ which are northern and southern hemispheres respectively. Sum of the azimuthal components of angular moments occupied by electrons in  $A$-partition is $L_z^A$ and $\xi$ represents the entanglement energy in an arbitrary unit.
		(b) Same as (a) but for an unequal number of electrons ($N_A=5$ and $N_B=9$) in two partitions.
		(c) Same as (a) but for $ \nu = 2/5 $ state with flux $ N_\Phi = 29 $ and $ N_A = N_B = 6 $.
		(d) Same as (c) but for an unequal number of particles in two partitions ($N_A=4$ and $N_B=8$).
	}
	\label{fig.es}
\end{figure}
We show (Fig.~\ref{fig.es}) the ES \cite{zozulya_2007_BipartiteEntanglementEntropy,haque_2007_EntanglementEntropyFermionic} for $\nu=1/2$ ($N=14$ particles with flux, $N_\Phi=27$) and $\nu=2/5$ ($N=12$ particles with flux, $N_\Phi=29$) in a moderate LLM strength of $\kappa =1.2$ belonging to the $\mathcal{A}$ phase. The low-lying spectra for these states in the $\mathcal{A}$ phase suggest that the sequence counting of edge states, which appears as 1-1-2-2-3-3-$\cdots$, is same for both the states. We have checked that this sequence of edge states is irrespective of the filling factors in the $\mathcal{A}$ phase.
This sequence of edge counting, as expected, is found to be independent of the total number of particles as well as the number of particles in a particular partition (hemisphere).

%% OVERLAP =======================================

\begin{figure}[]
	\centering
	\includegraphics[width=\linewidth]{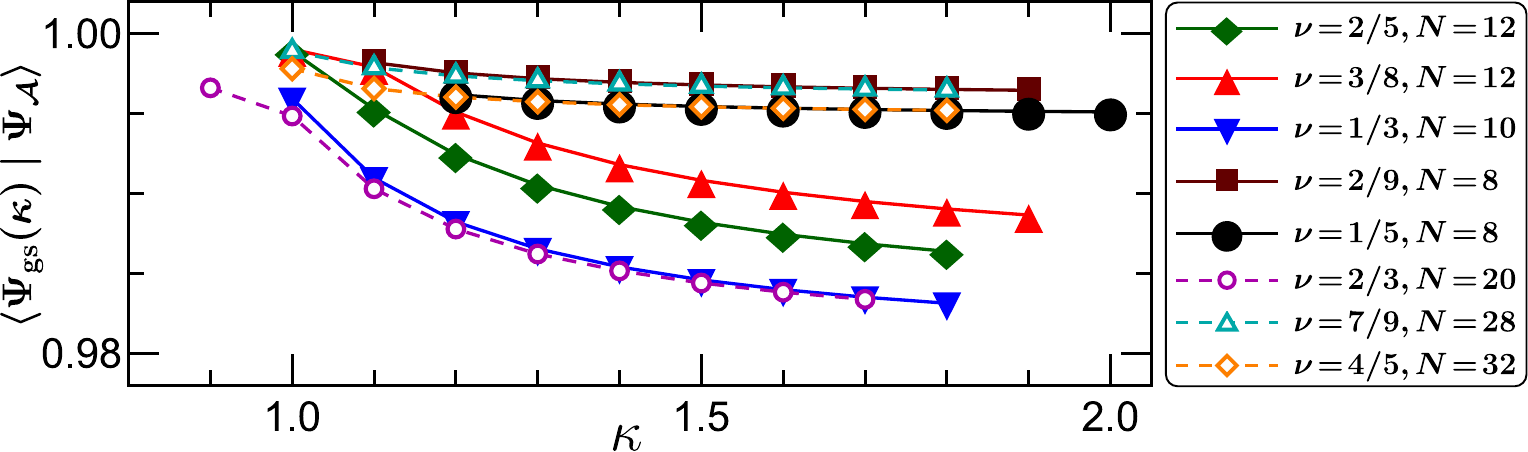}
	\caption{(color online) 
		Overlaps (with appropriate normalization) of the proposed wavefunctions, $\Psi_{\mathcal{A}}$ in Eq.~\eqref{eq.gen_wavefun}  with the exact ground states, $\Psi_{\text{gs}}(\kappa)$ of the Hamiltonian, $\hat{H}_{\text{eff}}(\kappa)$ in Eq.~\eqref{eq.llmpot} for noted $\nu$ and $N$ at different values of $\kappa$ in the corresponding regime of $\mathcal{A}$ phase.
	}
	\label{fig.overlap}
\end{figure}

Using the Monte-Carlo method with Metropolis algorithm (see Ref.~\onlinecite{jain_2007_CompositeFermions}), we calculate the overlaps of the exact ground states of $\nu = $ 2/5, 3/8, 1/3, 2/9, and 1/5 in the $\mathcal{A}$ phase with the proposed trial wavefunctions \eqref{eq.gen_wavefun} transformed into the form corresponding to the spherical geometry.
The overlaps for 2/3, 7/9, and 4/5 states are calculated \cite{overlap} with the particle-hole conjugate form of the exact ground states with the wavefunction in Eq.~\eqref{eq.gen_wavefun} for their respective conjugate filling factors 1/3, 2/9 and 1/5. 
All of our proposed wavefunctions in a general footing have remarkably high overlaps (Fig.~\ref{fig.overlap}) with the corresponding exact ground state wavefunctions for a range of $\kappa$.

The FQHE states such as 1/3 and 1/5 in the SLL are tagged as Abelian in literature \cite{dolev_2011_CharacterizingNeutralModes,wurstbauer_2015_GappedExcitationsUnconventional,macdonald_1986_CollectiveExcitationsFractional,dambrumenil_1988_FractionalQuantumHall,faugno_2021_UnconventionalMathbbZParton}.
However, the numerical studies \cite{macdonald_1986_CollectiveExcitationsFractional,dambrumenil_1988_FractionalQuantumHall,faugno_2021_UnconventionalMathbbZParton} are based on either zero or very low $\kappa$ where Laughlin wave functions have good overlaps at their respective flux shifts of $3N-3$ and $5N-5$. We find that these wavefunctions have negligible overlaps in the  $\mathcal{A}$ phase at their respective fluxes \cite{supplementary}. 
In contrary, these two states are also predicted to be non-Abelian characterized by the wave function $\Psi_{\mathcal{A}}$ in Eq.~\eqref{eq.gen_wavefun} in the $\mathcal{A}$ phase.

The predicted flux-shift here is 1 for all the FQHE states given by $\nu$ in Eq.~\eqref{eq.series} as well as its particle-hole conjugate filling factor $1-\nu$. The only previously proposed wavefunctions that have the same flux-shift that of ours are for 1/2 and 2/5 in Ref.~\onlinecite{jolicoeur_2007_NonAbelianStatesNegative} and 3/8 states in Refs.~\onlinecite{jolicoeur_2007_NonAbelianStatesNegative,bonderson_2008_FractionalQuantumHall}. However, the wavefunctions in Ref.~\onlinecite{jolicoeur_2007_NonAbelianStatesNegative} are not convenient for implementing in numerical comparison with any other wavefunctions.
The vanishingly small overlap of earlier proposed wavefunctions \cite{bonderson_2008_FractionalQuantumHall,hutasoit_2017_EnigmaNuFractional} for 3/8 with that of ours, suggests they belong to different topological classes, despite having the same flux-shift. In general, all the known trial wavefunctions which are reasonable descriptions \cite{moore_1991_NonabelionsFractionalQuantum,lee_2007_ParticleHoleSymmetryQuantum,levin_2007_ParticleHoleSymmetryPfaffian,son_2015_CompositeFermionDirac,zucker_2016_StabilizationParticleHolePfaffian,macdonald_1986_CollectiveExcitationsFractional,faugno_2021_UnconventionalMathbbZParton,read_1999_PairedQuantumHall,jolicoeur_2007_NonAbelianStatesNegative,bishara_2008_QuantumHallStates,bonderson_2008_FractionalQuantumHall,jain_1989_IncompressibleQuantumHall,balram_2018_FractionalQuantumHall,balram_2019_PartonConstructionParticleholeconjugate} of FQHE states near $\kappa \sim 0$ in the SLL  become irrelevant in the ${\mathcal A}$ phase.

We also show \cite{supplementary} the occurrence of FQHE states 4/11, 2/7, and 1/4 in the $\mathcal{A}$ phase of the SLL. Our proposed wavefunctions \eqref{eq.gen_wavefun} for these states have excellent overlap with the corresponding exact ground states. These FQHE states have not yet been observed. Experimental findings of these states will further confirm the validity of the general theory presented in this paper.

\begin{figure}[t]
	\centering
	\includegraphics[width=\linewidth]{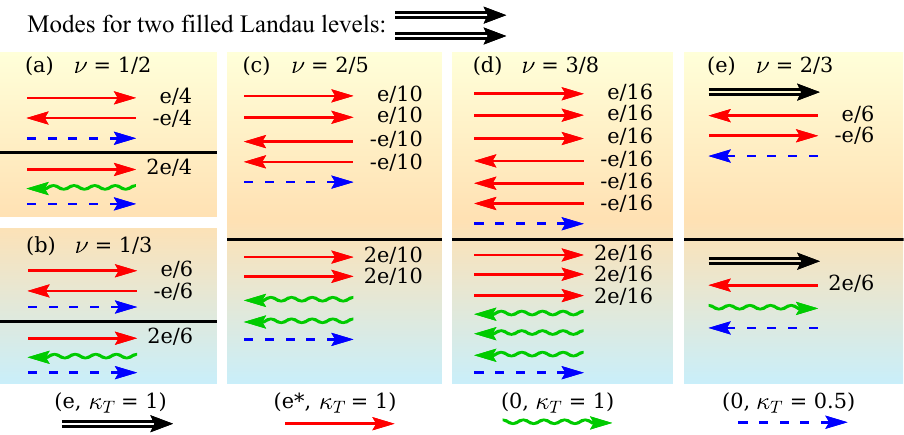}
	\caption{(color online) 
			Schematic of edge modes for different filling factors $\nu$ in the SLL. 
			Arrow-headed solid double-line, solid single line, wavy line, and dashed line, respectively, represent electronic mode, charged quasiparticle mode, charge neutral mode, and Majorana mode. 
			Each panel has two parts: (i) In the upper part, edge modes are shown as per the $\mathbb{K}$ matrix in Eq.~\eqref{eq.Kmatrix}; (ii) in the lower part, equivalent possibilities of modes due to disorder are shown. 
			Symbols $(e^\ast,\kappa_{_T})$ represent charge and thermal Hall-conductivity in the unit of $G_0$ carried by different modes respectively. The values of $e^\ast$ for quasiparticle/quasihole modes are shown beside the corresponding modes.
	}
	\label{fig.modes}
\end{figure}

%% K-Matrix =======================================
From the $ 2n $-component structure of the proposed wavefunction Eq.~\eqref{eq.gen_wavefun}, we extract the topological properties of the $ \mathcal{A} $ phase through the low-energy effective Lagrangian density \cite{wen_1992_ClassificationAbelianQuantum} as,
\begin{equation}\label{eq.lagrangian}
	\mathcal{L} = -\frac{1}{4\pi}\epsilon^{\alpha\beta\gamma} \sum_{I,J=1}^{2n} \mathbb{K}_{IJ} a_\alpha^I \partial_\beta a_\gamma^J - \frac{1}{2\pi}\epsilon^{\alpha\beta\gamma} \sum_{I=1}^{2n} t_I A_\alpha \partial_\beta a_\gamma^I
\end{equation}
Here $a_\alpha^I$  represents $I$-th component of $2n$-component Chern-Simons gauge fields, $A_\alpha$ is the external electromagnetic field, and $\epsilon^{\alpha\beta\gamma}$ is the antisymmetric Levi-Cevita tensor. The symmetric $ \mathbb{K} $-matrix for the sequence of states given in Eq.~\eqref{eq.series} can be read from the proposed wavefunction \eqref{eq.gen_wavefun} as
\begin{equation}\label{eq.Kmatrix}
	\mathbb{K} = 	\begin{pmatrix}
		\mathbb{C} & \mathbb{M} \\
		\mathbb{M} & \mathbb{C}
	\end{pmatrix} \,,
\end{equation}
where $ \mathbb{C} $ and $ \mathbb{M} $ are $n\times n$ matrices  given respectively by $ \mathbb{C}_{ij} = 1  $ and $ \mathbb{M}_{ij} = 2(m-1-\delta_{ij})+1$.
Further introducing charge vector $ t^T= (1,1,1, \cdots)_{2n} $ and quasiparticle vector $l^T = (1,0,0, \cdots)_{2n}$ and following Ref.~\onlinecite{wen_1992_ClassificationAbelianQuantum}, we find topological properties such as filling factor $\nu = t^T \mathbb{K}^{-1} t = n/(nm - 1)$ and quasiparticle charge $q= e\, l^T \mathbb{K}^{-1}t=e/[2(n m -1)]$.
Half of the eigenvalues of the $ \mathbb{K} $-matrix are positive, and the rest are negative.
Therefore, $n$ downstream and $n$ upstream quasiparticle/quasihole charge edge modes will exist for the FQHE states with $\nu=n/(nm-1)$ as well as its particle-hole conjugate filling factor $1-\nu$.
Owing to the presence of disorder, these quasihole charge modes may be converted \cite{kane_1994_RandomnessEdgeTheory,bid_2010_ObservationNeutralModes} into a neutral mode with an addition of an additional quasiparticle in the downstream mode.
Because $ \Psi_{\mathcal{A}}(n,m) $ in Eq.~\eqref{eq.gen_wavefun} indicates that every CB has choices of joining two available condensates which can accommodate $N/2$ of them and remains noninteracting, it has a hidden $ \mathbb{Z}_2 $ symmetry \cite{cappelli_2001_ParafermionHallStates}.
Consequently, there will be one neutral downstream (upstream) Majorana edge mode carrying $0.5G_0$ thermal Hall conductance for the sequence of filling factors $\nu$ ($1-\nu$). Figure \ref{fig.modes} illustrates the possible edge modes for $\nu=1/2$, 2/5, 1/3, 2/3, and 3/8 in the SLL. Considering two (three) completely filled Landau levels for $\nu$  ($1-\nu$) and thereby two (three) downstream bosonic edge modes, the total thermal Hall conductance will be
$2.5G_0$ irrespective of the FQHE states in the $\mathcal{A}$ phase of the SLL. This counterintuitive feature in thermal Hall conductance is, however, consistent with the identical counting of edge states in $\nu=1/2$ and $2/5$ shown in Fig.~\ref{fig.es}.

% @@@@@@@@@@@@@@@@@@@@@@@@@@@ SUMMARY @@@@@@@@@@@@@@@@@@@@@@@@@@@@@
% 
In summary, we find a sequence \eqref{eq.series} that exhausts \cite{note613} all (except 6/13 as argued)  the observed  FQHE states in the SLL. These states are shown to be incompressible in the $\mathcal{A}$ phase at the moderate regime of $\kappa \sim 1$. Our proposed trial wavefunctions \eqref{eq.gen_wavefun} for all these states have very high overlap with the corresponding exact ground states. The characteristics of these wavefunctions support non-Abelian quasiparticle excitations from their respective ground states.
Based on the proposed wavefunctions, we determine the possible edge modes and consequently predict $2.5G_0$ thermal Hall conductance for all these states.
Experimental confirmation on this prediction will place our theory as the most relevant one as no other theory has thus far predicted this unusual result.
Besides, we have also found \cite{supplementary} the signature of existence of the $\mathcal{A}$ phase in states like 4/11, 2/7, and 1/4 belonging to the proposed sequence in Eq.~\eqref{eq.series}. Observations of these states will further strengthen the validity of the proposed general theory of FQHE in the SLL.

% @@@@@@@@@@@@@@@@@@@@@@@@@  ACKNOWLEDGEMENTS @@@@@@@@@@@@@@@@@@@@@@@@@
%
We appreciate the generosity of the developers of the DiagHam package for keeping it open access. We acknowledge the Param Shakti (IIT Kharagpur) – a National Supercomputing Mission, Government of India, for providing computational resources.
S.S.M. is supported by the Science and Engineering Research Board (Government of India) through Grant No. MTR/2019/000546.

%\bibliography{references}% Produces the bibliography via BibTeX.

%merlin.mbs apsrev4-1.bst 2010-07-25 4.21a (PWD, AO, DPC) hacked
%Control: key (0)
%Control: author (8) initials jnrlst
%Control: editor formatted (1) identically to author
%Control: production of article title (-1) disabled
%Control: page (0) single
%Control: year (1) truncated
%Control: production of eprint (0) enabled
%

\clearpage

\renewcommand{\figurename}{FIG. S\!\!}
\renewcommand{\tablename}{TABLE S\!}
\renewcommand*{\thesection}{S\arabic{section}}
\renewcommand\theequation{S\arabic{equation}}
\renewcommand{\bibnumfmt}[1]{[S#1]}

\setcounter{equation}{0}
\setcounter{figure}{0}

%========================================================================================

\begin{widetext}
	\centering
	\Large{\bf{ Supplemental Material for ``Fractional Quantum Hall States of the $\mathcal{A}$ phase in the Second Landau Level"} }\\[1cm]
\end{widetext}

% @@@@@@@@@@@@@@@@@@@@@@@@@@@@ SUPPLEMENT @@@@@@@@@@@@@@@@@@@@@@@@@
% 
This supplemental material consists of three sections. In Section \ref{sec.gap}, we show the charge gap evaluated in spherical geometry using planner pseudopotential corrections up to three-body $V_9^{(3)}$ due to LLM together with planer bare Coulomb pseudopotentials.
In Section \ref{sec.extrastates}, we show some of the predicted FQHE states which are yet to be observed in the $\mathcal{A}$ phase. Existence of the $\mathcal{A}$ phase at the Laughlin flux for $\nu=1/3$ and $1/5$ is shown in Section \ref{sec.phaseLau}.

\begin{figure}[b]
	\centering
	\includegraphics[width=\linewidth]{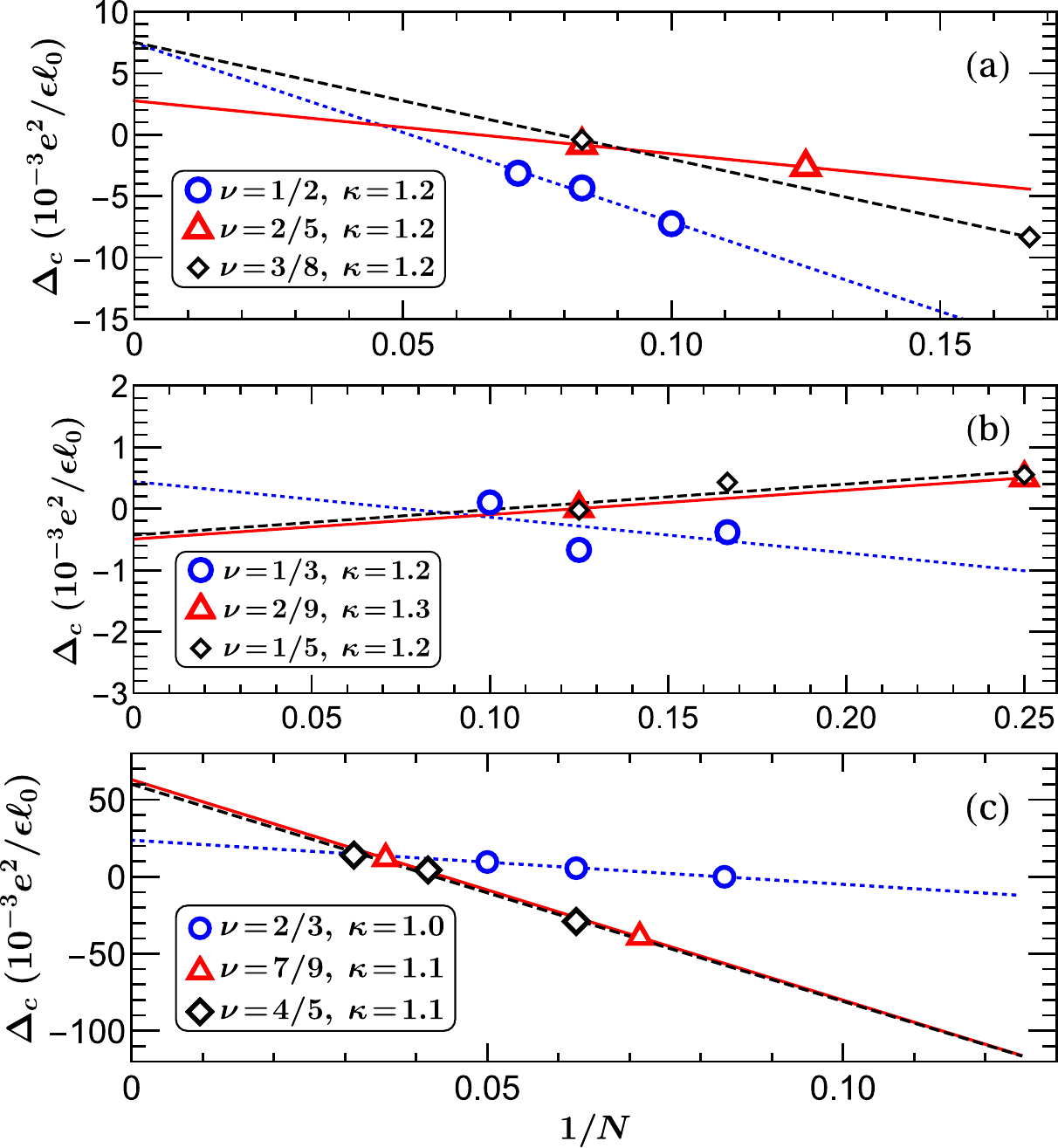}
	\caption{Charge gap for different filling factors scaled with $1/N$  as estimated using planner pseudopotentials  up to three-body $V_9^{(3)}$ together with two-body bare pseudopotentials and their corrections.}
	\label{fig.cgap}
\end{figure}

\section{\label{sec.gap}Charge Gap}
The charge gap is evaluated by taking the average of energies for creating a pair of quasiparticle and quasihole, i.e., $\Delta_c = (E_{\text{qp}}+E_{\text{qh}})/2$. Here, the quasiparticle energy $E_{\text{qp}}= E(N,N_\Phi-1)-E(N,N_\Phi)$ and quasihole energy $E_{\text{qh}} = E(N,N_\Phi+1)-E(N,N_\Phi)$ for an FQHE system given by $(N, N_\Phi)$ are estimated by subtracting the corresponding background energies, $ N^2/ \sqrt{2 N_\Phi} $ in the unit of $ e^2/(\epsilon \ell_0) $. Energies for finite systems are further corrected with the density correction factor $\sqrt{\nu N_\Phi /N}$.

The charge gaps are shown in Fig.~2 by considering the spherical bare Coulomb pseudopotentials and LLM corrections up to $ V_8^{(3)} $ pseudopotentials [54] estimated in the disk geometry. Fig.~S\ref{fig.cgap} shows charge gaps with disk three-body pseudopotentials up to $ V_9^{(3)} $ [31] including two-body bare Coulomb pseudopotentials and their corrections corresponding to disk geometry.

The thermodynamic charge gap remains finite and positive for all the fractional states regardless of the geometry chosen for bare pseudopotentials and inclusion of $ V_9^{(3)} $-pseudopotential except for $\nu=1/5$ and 2/9.
While the charge gaps found in Fig.~2 for these two states are small positive, these gaps turn out to be small negative in Fig.~S\ref{fig.cgap}. Therefore, a more accurate estimation of pseudopotential corrections (higher orders in $\kappa$) would be appropriate for making a conclusion about the occurrence of these two states in the $\mathcal{A}$ phase.

\begin{figure}[b]
	\centering
	\includegraphics[width=\linewidth]{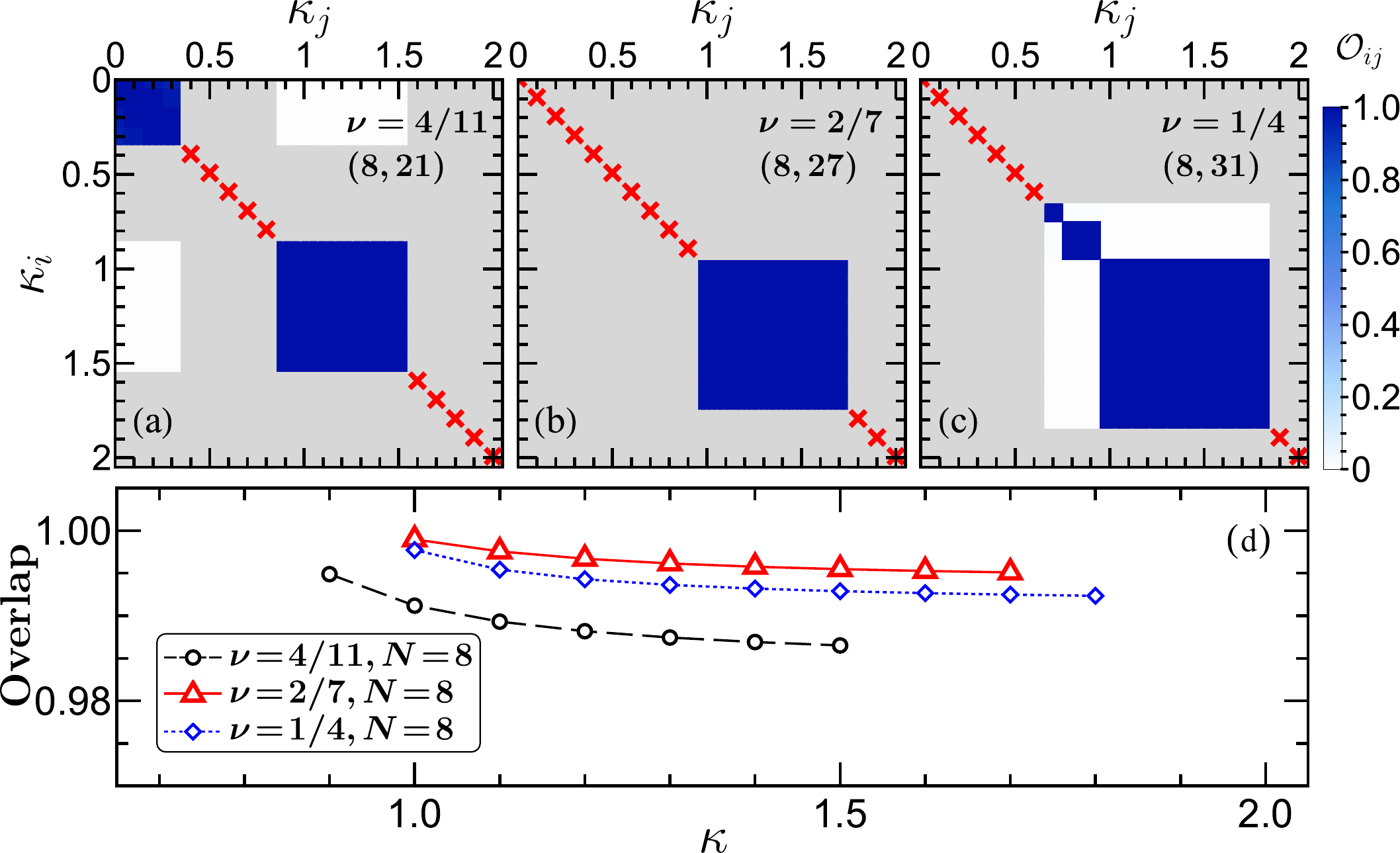}
	\caption{
		(a)--(c) Overlaps as described in the caption of Fig.~1 for $\nu=4/11,\,2/7$, and $1/4$.
		(d) Overlaps of the exact ground states with the proposed trial wavefunctions Eq.~(2) as described in the caption of Fig.~4 for $\nu=4/11,\,2/7$, and $1/4$.
	}
	\label{fig.extra}
\end{figure}

\section{\label{sec.extrastates}Predicted States in the SLL}
Apart from the observed states, the sequence $\nu = n/(nm-1)$ in Eq.~(1) shows many other possible states. However, those states must be analyzed energetically for the model Hamiltonian in Eq.~(3). In Fig.~S\ref{fig.extra}, we show  occurrence of the $\mathcal{A}$ phase for $\nu=4/11$ ($m=3$, $n=4$), 2/7 ($m=4$, $n=2$), and 1/4 ($m=5$, $n=1$) states as well.
The exact ground state wavefunctions for these states have excellent overlaps with the proposed wavefunction in Eq.~(2). These states have not been experimentally observed yet, but if found, it will further empower our proposed general theory for the SLL.

\section{\label{sec.phaseLau}Phase diagrams at Laughlin Flux}
\begin{figure}[h]
	\centering
	\includegraphics[width=\linewidth]{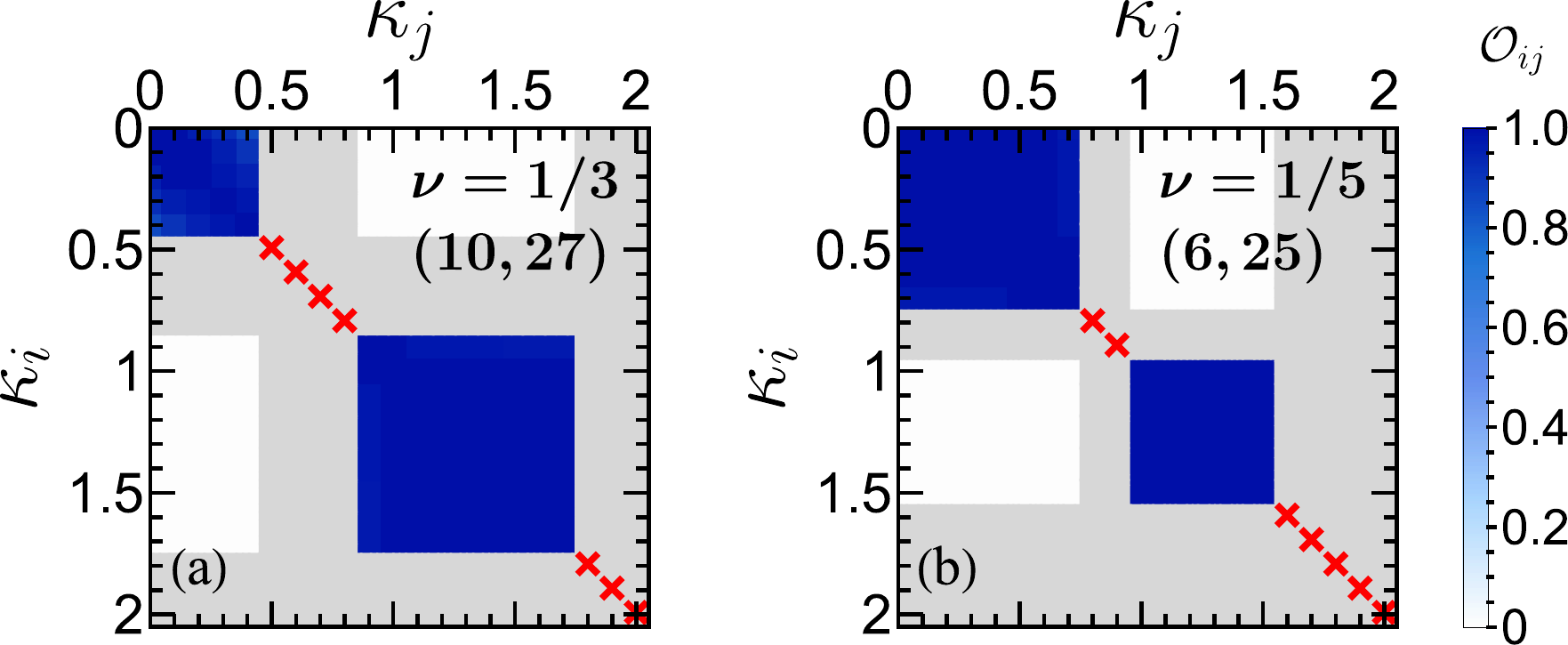}
	\caption{Same as Fig.~1 but for Laughlin flux shifts for 1/3 and 1/5 FQHE states in the SLL.}
	\label{fig.phaselau}
\end{figure}

The $\mathcal{A}$ phase is also found to be present for the FQHE states 1/3 and 1/5 at the Laughlin flux of $3N-3$ and $5N-5$, respectively, as shown in the Fig.~S\ref{fig.phaselau}.

%\bibliography{references}% Produces the bibliography via BibTeX.

\begin{thebibliography}{63}
\bibitem [{not()}]{note613}%
	\BibitemOpen
	\href@noop {} {}\bibinfo {note} {%Amongst all the observed FQHE states, the
		%filling factor $6/13$ is not studied here as it does not belong to the
		%sequence in Eq.~\eqref{eq.series} and the construction of its wavefunction
	%	will differ from Eq.~\eqref{eq.gen_wavefun}. 
	A detailed analysis of $6/13$
		state in $\mathcal{A}$ phase will be published elsewhere.}\BibitemShut
	{Stop}%
\bibitem [{ove()}]{overlap}%
	\BibitemOpen
	\href@noop {} {}\bibinfo {note} {%It is not convenient to write particle-hole
		%conjugate form of the wavefunctions in Eq.~\eqref{eq.gen_wavefun}. For
		%determining overlaps for 2/3, 7/9, and 4/5 states, we instead convert the
	%	exact ground state wavefunctions into their particle-hole conjugate states
	%	with the corresponding change in particle numbers followed by the
	%	determination of its overlap with the wavefunction in
	%	Eq.~\eqref{eq.gen_wavefun}. 
	For example, the exact wavefunction of 2/3 state
		for 12 particles has been converted into 1/3 state with 6
		particles.}\BibitemShut {Stop}%
\end{thebibliography}

\end{document}